\documentclass[aps, prl, onecolumn, showpacs, preprintnumbers, amsmath, amssymb]{revtex4}

\usepackage{graphicx, color}
\usepackage{dcolumn}
\usepackage{bm}

\usepackage{verbatim}

\begin{document}

 \author{M.~E.~Povarnitsyn$^{1}$, T.~E.~Itina$^2$, K.~V.~Khishchenko$^1$, and P.~R.~Levashov$^1$}
 \affiliation{$^{1}$Joint Institute for High Temperatures of RAS, Izhorskaya 13 Bldg 2, Moscow, 125412, Russia \\ $^{2}$Laboratoire Hubert Curien, UMR CNRS 5516, 18 rue Beno\^it Lauras, B\^at. F, 42000, St-Etienne, France}

\title{Suppression of ablation in femtosecond double pulse experiments}
\date{\today}

\begin{abstract}
We report the physical reasons of a curious decrease in the crater depth observed for long delays in experiments with femtosecond double  pulses. Detailed hydrodynamic modeling demonstrates that the ablation mechanism is dumped when the delay between the pulses exceeds the electron-ion relaxation time. In this case, the interaction of the second laser pulse with the expanding target material leads to the formation of the second shock wave suppressing the rarefaction wave created by the first pulse. The evidence of this effect follows from the pressure and density profiles obtained at different delays after the first laser pulse.
\end{abstract}

\pacs{52.38.Mf, 79.20.Ds, 42.25.Bs, 72.15.Cz}

\maketitle

During last decade, femtosecond laser systems have found numerous applications in different areas, such as laser material treatment, nano-optics, and surface analysis~\cite{Book:femto}. One of the particularly promising applications of ultrashort laser pulses is  laser-induced breakdown spectroscopy (LIBS)~\cite{LIBS:2006}, or a remote technique that  can be used to instantly analyze different materials. The advantages of ultrashort laser pulses were demonstrated in several LIBS experiments resulting in a considerably improved spectral resolution.  To further increase the accuracy of the analysis, several special configurations were proposed enhancing the intensity of the plume emission. One of the promising schemes is based on double pulse (DP) laser set-ups~\cite{Semerok:2004, Fotakis:2008, Noel:PhD}. In addition, experiments with several delayed femtosecond pulses were shown to be advantageous in laser machining~\cite{Stoian:2002}, thin film modification~\cite{ZHan:2008} and in fast ion production~\cite{Guillermin:2009}.

\begin{figure}
\begin{center}
\includegraphics[width=0.95\columnwidth]{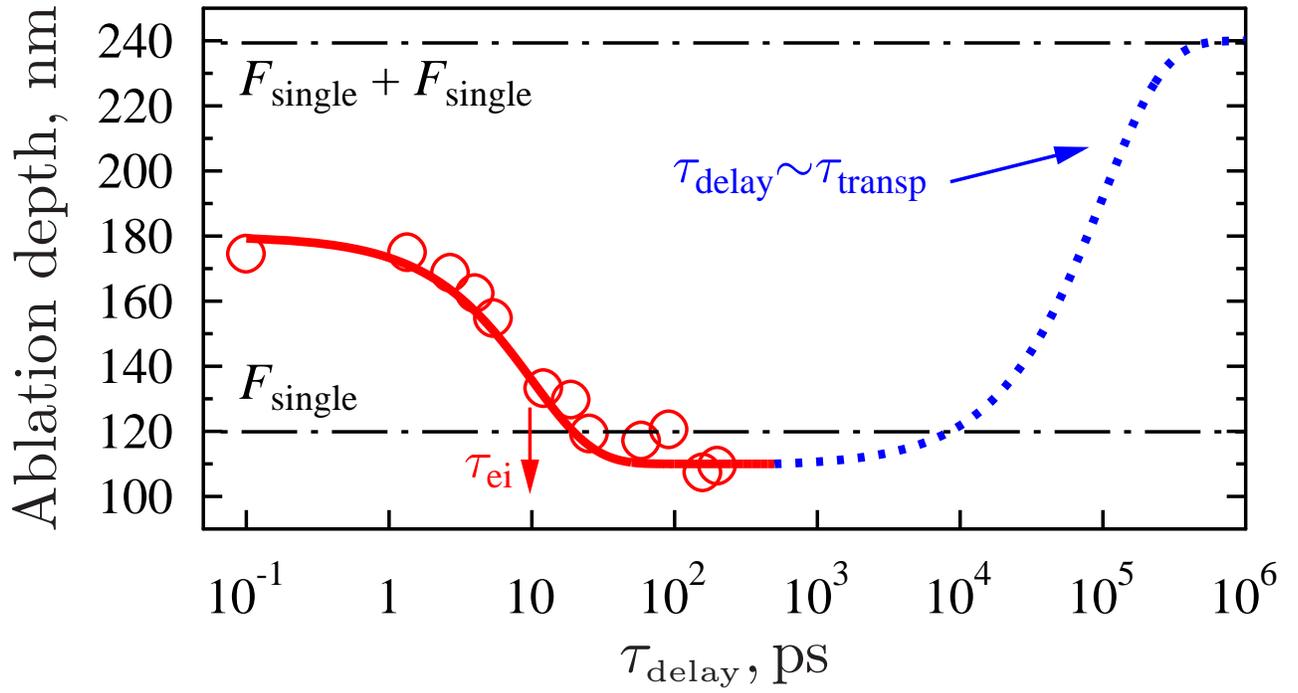}
\end{center}
\caption{(Color online) Empty (red) circles---DP experiment~\cite{Noel:PhD} with Cu ablation by pulses with $F_\textnormal{single}=2$~J/cm$^2$; solid (red) curve---interpolation of the experiment. Dashed (blue) curve---our theoretical assumption for $\tau_\textnormal{delay}\rightarrow\infty$. Time $\tau_\textnormal{transp}$ is the moment of the ablated substance transparency for the second pulse due to rarefaction. Horizontal dash-and-dot (black) lines show depths from SP and two pulses separated by  $\tau_\textnormal{delay}\rightarrow\infty$.} \label{fig1}
\end{figure}

Performing DP femtosecond experiments for different metals, several authors~\cite{Semerok:2004, Noel:PhD, Donnelly:2009} surprisingly noticed a monotonic decrease in the resulting crater depth  with the increase in the delay $\tau_\textnormal{delay}$  between the laser pulses. This effect was observed both under vacuum~\cite{Noel:PhD, Donnelly:2009} and in air~\cite{Semerok:2004}. In the experiments~\cite{Noel:PhD} with copper targets, for instance, the laser fluence of each pulse was set to be $F_\textnormal{single}= 2$~J/cm$^2$. When the delay between the pulses was much shorter than the electron-ion relaxation time $\tau_{ei}$ in the target material $(\tau_\textnormal{delay}\ll\tau_{ei}$, where $\tau_{ei}\approx10$~ps for copper~\cite{Handbook:chemistry:1999}) the crater depth was the same as in the case of a single pulse (SP) with the laser fluence $2\times F_\textnormal{single}=4$~J/cm$^2$. For the delays approaching the electron-ion relaxation time $(\tau_\textnormal{delay} \sim \tau_{ei})$, the crater depth monotonically decreased. Finally, for longer delays $(\tau_\textnormal{delay}\gg\tau_{ei})$  the crater depth was found to be even smaller than that obtained with a SP at the fluence $F_\textnormal{single}$, see Fig.~\ref{fig1}. Similar behavior was also obtained for aluminum~\cite{Semerok:2004}, gold~\cite{Noel:PhD} and nickel~\cite{Donnelly:2009} thus indicating the common physical effect. The dashed (blue) curve in Fig.~\ref{fig1} is just a prediction for long delays, which have not been yet achieved in the DP experiments. This case will be discussed at the end of the letter.

Recently, several explanations of the unusual dependency of the crater depth on the delay were proposed. In particular, the temperature dependence of the heat conductivity claimed to explain the observed effect in~\cite{Noel:PhD}.  Target heating can in fact lead to the energy accumulation changing target absorption properties. In addition, the energy absorption in the laser-generated plasma plume was furthermore evoked in~\cite{Semerok:2004}. The latter explanation makes sense if the time delay is sufficiently long, so that the second pulse interacts with the laser-created expanding plasma plume. Recently~\cite{Donnelly:2009},the ablation reduction effect was also attributed to the presence of a liquid layer~\cite{Perez:2009}.  In fact, laser ablation of metals is commonly accompanied by melting. The supercritical concentration of the conduction electrons in liquid phase makes, however, the melted layer non-transparent for laser light. Therefore, a better understanding of the experimental findings~\cite{Semerok:2004, Noel:PhD, Donnelly:2009} requires a more complex self-consistent analysis of an interplay of different physical effects to highlight the basic mechanism of suppression of ablation. A detailed hydrodynamic modeling can provide a comprehensive physical scenario of the interaction dynamics.  In the present letter, we report the results of such numerical calculations based on a two-temperature hydrodynamic model. The model accounts for such processes as laser light absorption, plume expansion, thermal conductivity, two-temperature effects, phase transitions, material decomposition, etc. In particular, the dynamics of the DP ablation is investigated in details and is shown to play a crucial role in the considered phenomenon.

 The hydrodynamic model~\cite{Povarnitsyn:ASS:2009, Povarnitsyn:PRB:2007} was developed and previously used for the investigation of a SP ablation of metals. The main advantages of the model are in the treatment of metastable phases and in the used multiphase wide-range equation of state~\cite{Povarnitsyn:ASS:2009}. The validity of the model was confirmed by the comparison with the experimentally-obtained ablation depth for several metals~\cite{Semerok:2001}. In the case of a DP, the model accounts for the laser energy absorption by the ablation plume generated by the first pulse. To treat the absorption process, we use the Helmholtz wave equation with the quasi-classic solution described elsewhere~\cite{Andreev:Elbrus2009:EB}. The absorbed laser power density (per unit volume) is expressed in this case as $Q_{L}=I_0k_0\textnormal{Im}\left\{\varepsilon\right\}\left|{E}/{E_0}\right|^2.$ Here $I_0$ is the peak laser intensity, $k_0=\omega_L/c$ is the laser wave vector with the laser frequency  $\omega_L$ and the light speed in vacuum $c $, $\varepsilon$ is the complex permittivity of media, $E(t, x)$ is the solution to the Helmholtz equation and $E_0=\sqrt{8\pi I_0/c}$ is the maximum of laser electric field in vacuum. The wide-range model of permittivity~\cite{Veysman:JEPT:2007} describes Drude-like (intraband) effects, band-to-band transitions and the hot plasma limit. The electron thermal conductivity in metals can be measured experimentally only for low and near-room temperatures~\cite{Handbook:metals:1998}. To extend the model~\cite{Eidmann:2000} used in this work over the range of high temperatures we apply the first-principle data~\cite{Redmer:2000, Vanina:2005}. The model for electron-ion coupling (electron-phonon for low temperatures) is similar to one used in~\cite{Eidmann:2000} and is also adjusted to satisfy the VASP (\textit{Vienna Ab-initio Simulation Package}) first-principle calculations~\cite{Zhigilei:2008}.

\begin{figure}
\begin{center}
\includegraphics[width=0.95\columnwidth]{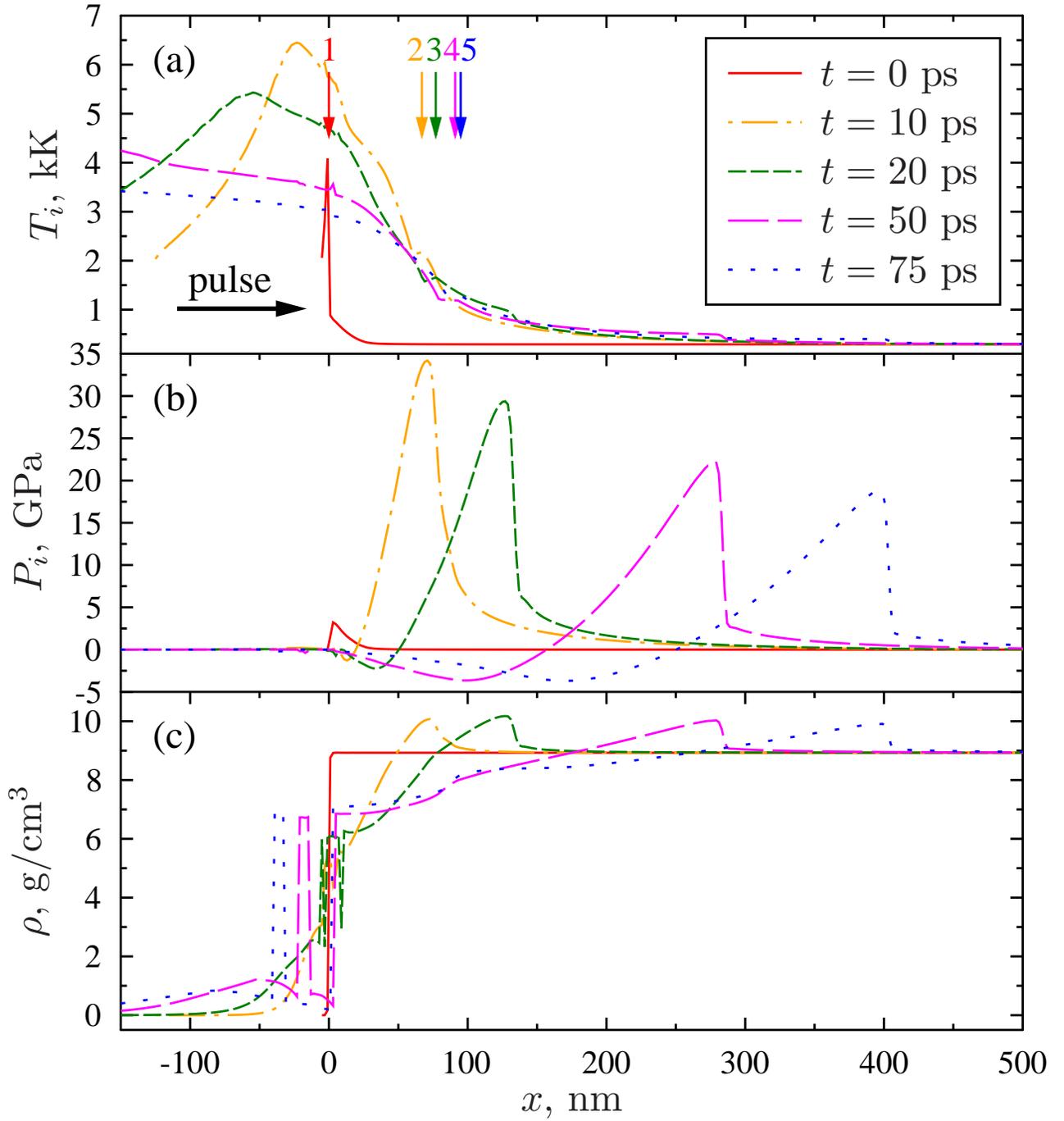}
\end{center}
\caption{(Color online) Calculation results for Cu target ablation in vacuum by a SP with fluence 2~J/cm$^2$.
Temperature---(a), pressure---(b) and material density ---(c) of the ions are shown for different moments. The initial free surface of the target is located at $x=0$~nm, the laser pulse goes from the left. Arrows show the melting front position $x_\textnormal{melt}$ for different times (1---0~ps, 2---10~ps, 3---20~ps, 4---50~ps, 5---75~ps).}
\label{fig2}
\end{figure}

Calculations are performed for copper with two $s$-polarized Gaussian laser pulses of 100~fs width, 800~nm wavelength, and with first and second pulse intensity maximum at $t=0$ and $t=\tau_\textnormal{delay}$, respectively.  Fig.~\ref{fig2} shows the dynamics of both shock and rarefaction waves in the copper target during and after the propagation of the first pulse. At $t=0$~ps, the ion pressure increases due to the collisions with hot electrons.  After the electron-ion collisional relaxation, the temperatures of the electron and ion sub-systems are the same, but the pressure of ions is much larger (up to 35~GPa at 10~ps in Fig.~\ref{fig2}b). At the moment of 20~ps after the peak of the first pulse, one can see the appearance of the negative pressure zone (with pressure about $-2.5$~GPa at $x\approx35$~nm). Later, at moments of 50~ps and 75~ps after the maximum of the first pulse, the negative pressure value is about $-3.6$~GPa located at $x\approx100$ and $x\approx175$~nm depth, respectively. Such pressure values were shown to result in the mechanical spallation in the liquid phase, but are insufficient for the spallation in the solid phase. The arrows in Fig.~\ref{fig2}a show the position of the melting front $x_\textnormal{melt}(t)$ for the corresponding moments. It is seen that the melting front propagation speed drops as the laser energy dissipates. For $x>x_\textnormal{melt}$ the substance is in solid phase and the spallation strength for a solid material is close to its theoretical limit~\cite{Zhigilei:2004}.

Our previous calculations~\cite{Povarnitsyn:PRB:2007, Povarnitsyn:ASS:2009} demonstrated that only the melted region can be ablated. Therefore, if the second pulse arrives during the rarefaction wave propagation (10~ps~$<t<$~50~ps in Fig.~\ref{fig2}a, b) through the liquid layer, this pulse can reheat the nascent ablation plume. As a result, a high-pressure region is generated in the vicinity of the initial target surface. This hot region then produces a shock wave and thus diminishes the action of the rarefaction wave.

\begin{figure}
\begin{center}
\includegraphics[width=0.95\columnwidth]{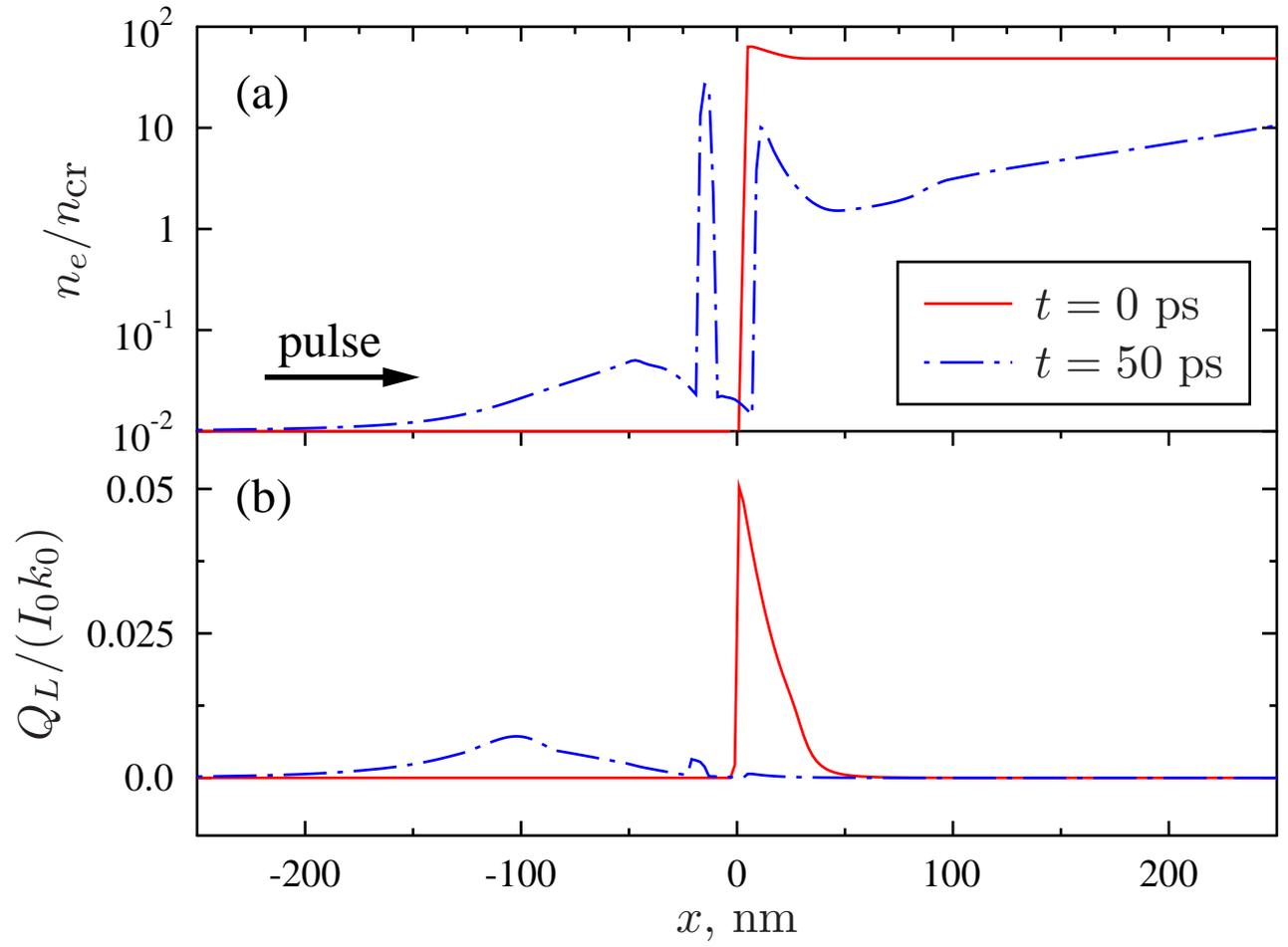}
\end{center}
\caption{(Color online) DP with $\tau_\textnormal{delay}=50$~ps for Cu target. The ratio of the electron concentration to the critical one---(a), relative absorption efficiency---(b). The first pulse maximum is at $t=0$~ps, the second one is at $t=50$~ps. Initial target position is at $x=0$~nm.}
\label{fig3}
\end{figure}

Fig.~\ref{fig3} illustrates the location of the laser energy absorption for both the first and the second pulses.  The ratio of
the free electron concentration $n_e$ to the critical one $n_\textnormal{cr}=\omega_L^2 m_e/(4\pi e^2)$ is presented in
Fig.~\ref{fig3}a. When the ratio $n_e/n_\textnormal{cr}\geq 1$, one can observe the common skin effect for the first pulse. For the second pulse, the electron concentration gradient is smooth and the absorption region is shifted to the left of the initial free surface position $x=0$~nm. The spike in the electron concentration in Fig.~\ref{fig3}a at 50~ps correlates with the broken off chunk seen in the density profile in Fig.~\ref{fig2}c for the same moment. Thus, the absorption  of the first pulse takes place from 0 to 50~nm in the skin layer (solid (red) curve), whereas the second pulse is absorbed from $-200$~to 0~nm (dash-doted (blue) curve) with the maximum at about $-100$~nm ($n_e\sim 0.03n_\textnormal{cr}$).

\begin{figure}
\begin{center}
\includegraphics[width=0.85\columnwidth]{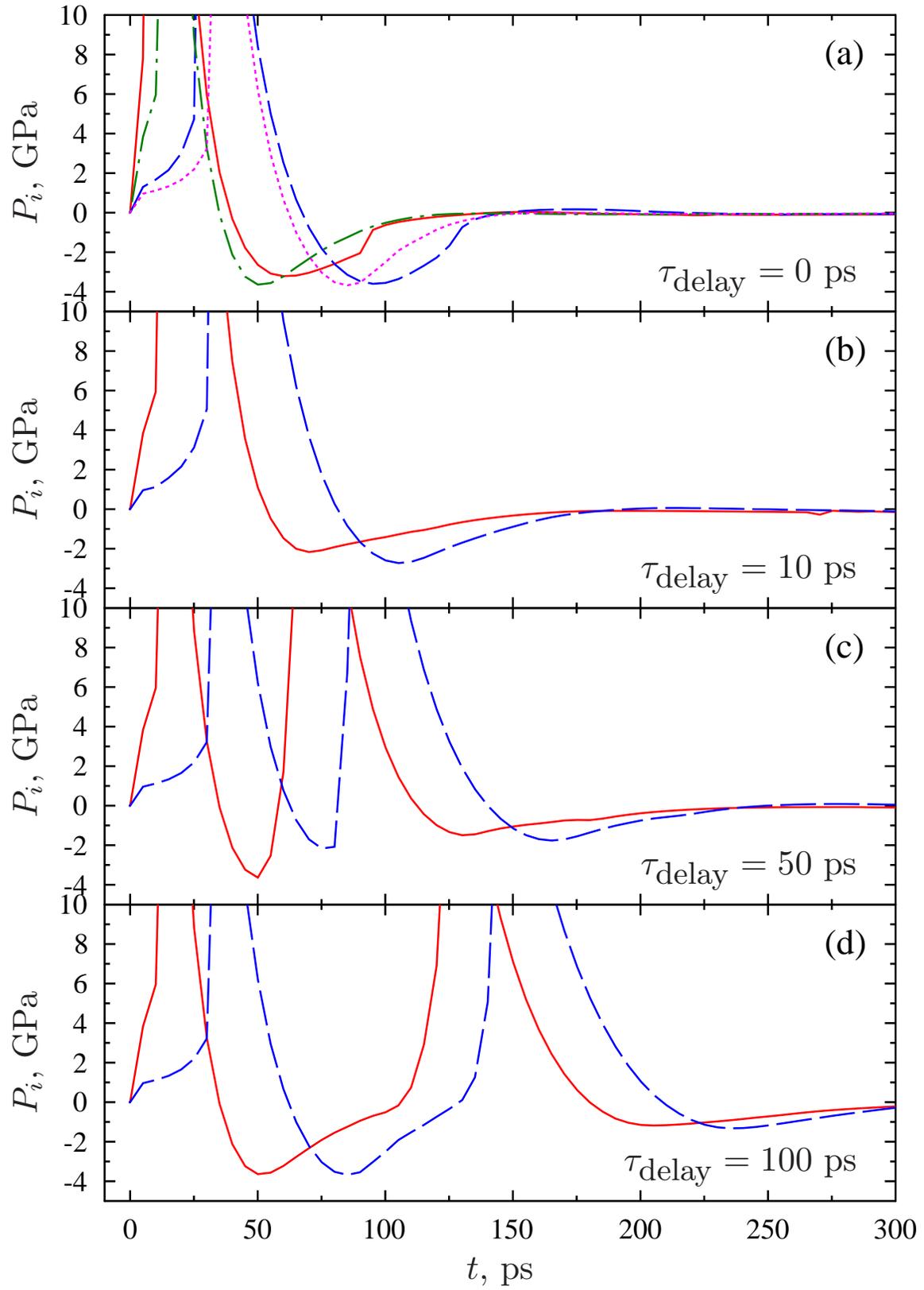}
\end{center}
\caption{(Color online) Evolution of the ion pressure for 100 and 200~nm depth in Cu target. Results presented for different delays between femtosecond pulses $2\times2$~J/cm$^2$: solid (red) curve---100~nm depth, long dashed (blue) curve---200~nm depth. SP pressure profiles for 2~J/cm$^2$ are also presented: dash-and-dot (green) curve---100~nm depth, short dashed (magenta) curve---200~nm depth.}
\label{fig4}
\end{figure}

To clearly illustrate the suppression of the rarefaction wave in the DP experiments, Fig.~\ref{fig4} presents the pressure evolutions calculated at the depth of 100 and 200~nm under the initial surface position. For the delay
$\tau_\textnormal{delay}=0$~ps (Fig.~\ref{fig4}a), the amplitude of the rarefaction wave is about $-3.5$~GPa both in the calculations with SP and DP. The melted layer can, therefore, be ablated forming a crater. At the delay of 5~ps, the rarefaction wave
is formed later, and the amplitude of the wave decreases ($\sim-3$~GPa). For longer delays, the minimal pressure
appears also later (at 200~nm depth and for 10 (Fig.~\ref{fig4}b) and 25~ps delay the moments are 100 and 125~ps, respectively), whereas the
amplitude of the negative pressure decreases (at 200~nm depth for 10 and 25~ps delay the pressure values are $-2.8$ and
$-2.3$~GPa, respectively). The observed attenuation of the pressure amplitude results in the decrease of the crater depth. This effect is observed despite the fact that in our simulations the the melted layer is thicker for longer delays between pulses. Finally, for the delays of 50 and 100~ps (Fig.~\ref{fig4}c, d), the first pulse leads to the spallation of the melted layer, whereas the intensity of the second rarefaction waves (at about 150 and 200~ps, respectively) is about $-1.8$~GPa. In this case, the ablated depth is close to that obtained for the SP with 2~J/cm$^2$.

\begin{figure}
\begin{center}
\includegraphics[width=0.95\columnwidth]{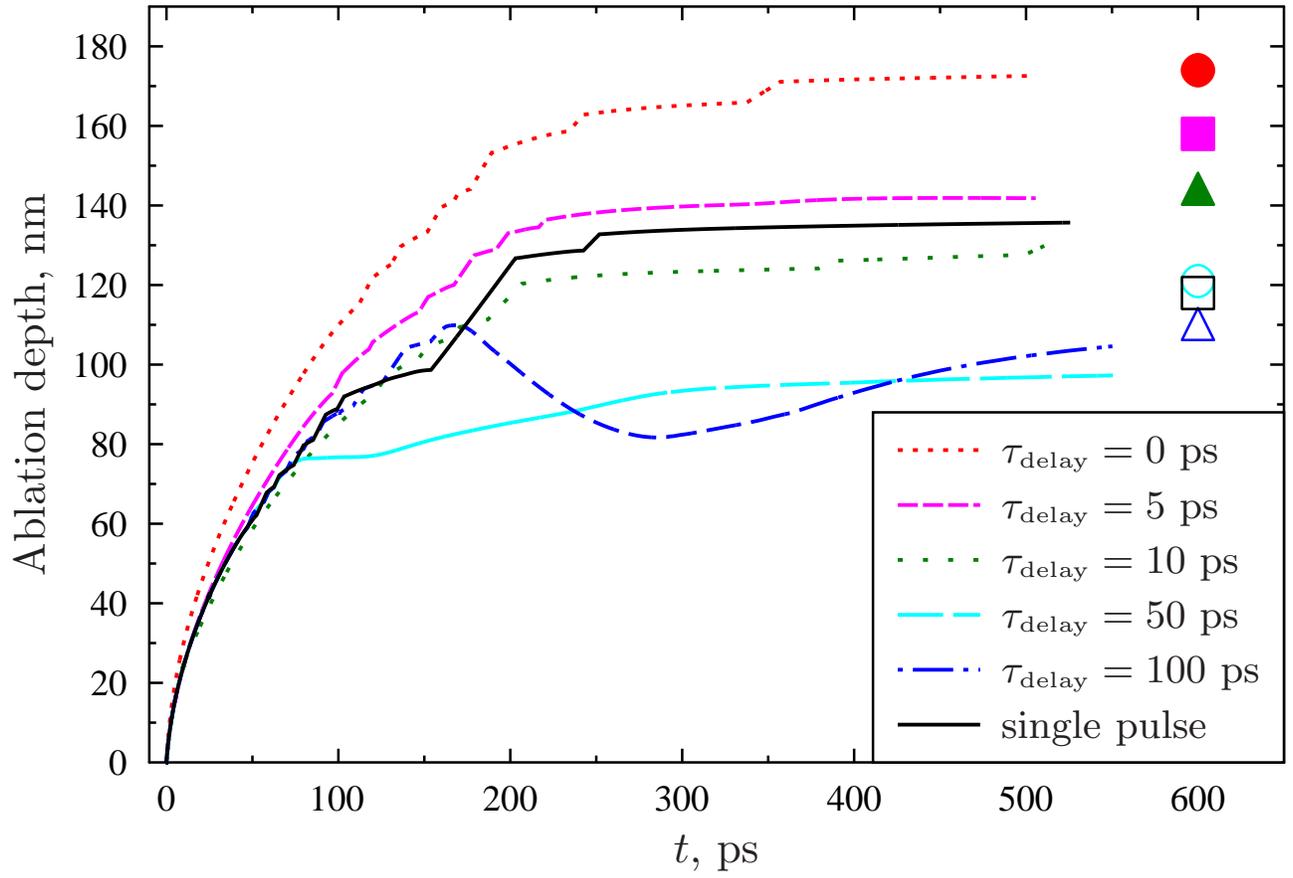}
\end{center}
\caption{(Color online) Ablation of Cu target. Simulation of crater depth $\Delta_\textnormal{abl}$ growth in time for different DP delays. Crated depth from experiment~\cite{Noel:PhD} for different delays is also presented: filled (red) circle---0~ps, filled (magenta) square---5~ps, filled (green) triangle---10~ps, empty (cyan) circle---50~ps, empty (blue) triangle---100~ps, empty (black) square---SP.}
\label{fig5}
\end{figure}

The performed simulations give us an opportunity to observe the dynamics of the ablation process. We suppose here that the crater depth is formed due to the material removing after the laser irradiation. We can calculate the ablation depth $\Delta_\textnormal{abl}$ integrating the mass flux through the plane $x=0$ (initial target surface) by using the equation $\Delta_\textnormal{abl}(t)=\rho_0^{-1}\int_0^t(\rho u)|_{x=0}dt'$, where $\rho_0$ is the initial material density and $u$ is the material speed. As one can see in Fig.~\ref{fig5}, the removed mass drops while the delay between pulses increases. Already for the 10~ps delay the crater depth for a DP case is equal to that obtained for a SP (short dotted (green) curve and solid (black) curve in Fig.~\ref{fig5}, respectively). For longer delays (50 and 100~ps) the crater depth is even smaller than in the case of a SP. The absorption of the second pulse in the nascent ablation plume results in the reheating of ablated material and acceleration of outward part of the plume and deceleration of the inner part of the plume. One can see in Fig.~\ref{fig5} ($\tau_\textnormal{delay}=100$~ps) that the back flux dominates from 170 to 280~ps and results in temporal decrease in the crater depth for 100~ps delay. The corresponding experimental points are presented to prove the validity of the model.

We can also suggest the crater formation mechanism in the case $\tau_\textnormal{delay}\rightarrow\infty$ (Fig.~\ref{fig1}, dashed (blue) curve). During the plume expansion, the electron concentration drops and the plasma cloud becomes transparent for the second pulse. This effect can be achieved for $t_\textnormal{transp}\approx R / u$ with the radius of the plume $R$ and the rarefaction velocity $u$. Simple estimations yield $t_\textnormal{transp}$ on the order of hundreds of nanosecond.

In summary, we have revealed the main stages of the DP ablation process. The first pulse is absorbed by the conduction band electrons in metal leading to a rapid rise in their temperature and pressure. During the electron-ion relaxation, the temperature of ions reaches the melting point, so that homogeneous melting occurs. At this time ($t\sim \tau_{ei}$), the pressure of ions rapidly grows resulting in the formation of a shock wave propagating into the target. At the same time, a rarefaction wave is formed near the free surface. This rarefaction wave propagates through the melted surface layer and gives rise to the mechanical fragmentation and ablation of the liquid target material. In the DP experiments, the delay between laser pulses plays a fundamental role. When the delay is much shorter than the relaxation time, only one shock wave and one rarefaction wave appear. In this case, the ablation crater is formed by both pulses simultaneously as in the case of a SP of the same energy. When the delay is on the order of $\tau_{ei}$, the second pulse reduces the intensity of the rarefaction wave and the depth of the ablation crater decreases. Finally for long delays $\tau_\textnormal{delay}\gg \tau_\textnormal{ei}$ the ablation crater is formed by the first pulse only, whereas the second pulse reheats the ablated material and does not cause any additional fragmentation of the target. We conclude that the demonstrated effect of the suppression of the rarefaction wave by the second laser pulse explains clearly the results of DP laser experiments.

Acknowledgments. We gratefully acknowledge support from the French-Russian collaboration project (CNRS/ASRF) 21276, and the Russian Foundation for Basic Research (grants 08-08-01055 and 09-08-01129). We wish to thank Dr. J\"org Hermann (LP3 CNRS, Marseille, France) for the discussions of DP laser experiments.




\end{document}